# Interfacial properties of 2D $WS_2$ on $SiO_2$ substrate from x-ray photoelectron spectroscopy and first-principles calculations

Changjie Zhou[1], Huili Zhu[1,†], Weifeng Yang[2,‡], Qiubao Lin[1], Tongchang Zheng[1], Lan Yang[1], Shuqiong Lan[1]

[1] Xiamen Key Laboratory of Ultra-Wide Bandgap Semiconductor Materials and Devices, Department of Physics, School of Science, Jimei University, Xiamen 361021, China

[2] Department of Microelectronics and Integrated Circuits, Xiamen University, Xiamen 361005, China

Corresponding authors. E-mail: [†] hlzhu@jmu.edu.cn; [‡] yangwf@xmu.edu.cn

**Abstract**

Two-dimensional (2D) WS$_2$ films were deposited on SiO$_2$ wafers, and the related interfacial properties were investigated by high-resolution x-ray photoelectron spectroscopy (XPS) and first-principles calculations. Using the direct (indirect) method, the valence band offset (VBO) at monolayer WS$_2$/SiO$_2$ interface was found to be 3.97 eV (3.86 eV), and the conduction band offset (CBO) was 2.70 eV (2.81 eV). Furthermore, the VBO (CBO) at bulk WS$_2$/SiO$_2$ interface is found to be about 0.48 eV (0.33 eV) larger due to the interlayer orbital coupling and splitting of valence and conduction band edges. Therefore, the WS$_2$/SiO$_2$ heterostructure has a Type I energy-band alignment. The band offsets obtained experimentally and theoretically are consistent except the narrower theoretical bandgap of SiO$_2$. The theoretical calculations further reveal a binding energy of 75 meV per S atom and the totally separated partial density of states, indicating a weak interaction and negligible Fermi level pinning effect between WS$_2$ monolayer and SiO$_2$ surface. Our combined experimental and theoretical results provide proof of the sufficient VBOs and CBOs and weak interaction in 2D WS$_2$/SiO$_2$ heterostructures.

# 1 Introduction

Tungsten disulfide ($WS_2$), one of two dimensional (2D) layered transition metal dichalcogenides (TMDCs), has been extensively investigated for its extraordinary mechanical, electronic and optoelectronic properties [1-12]. Structurally, $WS_2$ is a typical hexagonal layered compound with strong in-plane and weak out-of-plane interaction, resulting in the absence of dangling bonds on the surface [13-15]. Unlike gapless graphene, 2D layered $WS_2$ exhibits a moderate and variable bandgap (1-2 eV). When cleaved down to monolayer (1L), energy bandgap of layered $WS_2$ converts from indirect to direct due to the quantum confinement effect [16,17]. In addition, $WS_2$ monolayer presents strong valley-contrasting spin splitting near the valence and conduction band edges owing to the absence of inversion symmetry and strong spin-orbit coupling [18]. These intriguing properties make 2D $WS_2$ not only a testing platform for various physical phenomena (valley polarization [19], valley Hall effect [20], and valley Zeeman effects [21], *etc*.), but also a promising material in a wide range of applications, such as field-effect transistors (FETs) [22,23], flexible electronics and optoelectronics [24-26], and various types of sensors [23,27-30].

Heterostructures of 2D layered $WS_2$ with insulating substrates are of fundamental importance in the electronic and optoelectronic applications [31,32]. Among various insulating substrates, silicon dioxide ($SiO_2$) is the most widely used one. On one hand, high-quality 2D $WS_2$ films were almost grown or exfoliated on the $SiO_2$ substrates. For example, $WS_2$ films with 400 μm single crystal domain have been fabricated on $SiO_2$ wafers by using atmospheric pressure chemical vapor deposition (CVD) [6]. Moreover, many previous works have demonstrated the high-performance $WS_2$-based FETs and photodetectors on $SiO_2$ substrates [22,23]. Iqbal *et al.* have successfully tailored the electronic and optoelectronic properties of a $WS_2$-based FET by using the selective chemical doping [22]. An unprecedented high on/off ratio of $10^8$ and a high-mobility of 255 $cm^2V^{-1}s^{-1}$ at room temperature were obtained in their work [22]. Huo *et al.* demonstrated that the photoresponsivity ($R_\lambda$) and external quantum efficiency (EQE) of FET based on the $WS_2/SiO_2$ heterostructure could be significantly enhanced to 884 $AW^{-1}$ and $1.7 \times 10^5$%, respectively, due to the inclusion of molecular $NH_3$ [23]. Most recently, the $WS_2/SiO_2$ heterostructures were also adopted to fabricate the non-volatile flash memory devices, where the layered $WS_2$ films were used as charge trapping stack layers [33].

In terms of testing the fundamental and novel phenomena of 2D layered $WS_2$, superconductivity in CVD-grown monolayer $WS_2$ is achieved on the $SiO_2$ substrate by strong electrostatic electron doping of an electric double-layer transistor [34]. By adopting the bilayer $WS_2/SiO_2$ heterostructure, Nayak *et al.* have achieved a record-high valley polarization of 80% at room temperature [19]. Strong valley Zeeman effect of dark excitons in monolayer $WS_2/SiO_2$ heterostructure has also been demonstrated in a tilted magnetic field [21]. All above reports revealed the importance and suitability of $SiO_2$ as the insulating substrate for the promising application of 2D $WS_2$. Then, Ulstrup *et al.* used photoemission electron microscopy (PEEM) to investigate the real- and momentum-space electronic structures of electrically contacted monolayer WS$_2$ stacked on SiO$_2$ substrates [35]. However, the interfacial properties of the WS$_2$/SiO$_2$ heterostructure, especially the atomic-scale insight into the interfacial interactions and band offsets at the interface, remain poorly understood.

In this study, the interfacial properties of 2D layered WS$_2$ grown on SiO$_2$/Si wafers were studied by x-ray photoelectron spectroscopy (XPS) and first-principles calculations. By using the experimental direct (indirect) method, the VBO value of 3.97 eV (3.86 eV) was obtained at 1L WS$_2$/SiO$_2$ interface, while the CBO was obtained to be 2.70 eV (2.81 eV). When increasing the WS$_2$ thickness to its bulk limit, the VBO and CBO were enhanced by 0.48 and 0.33 eV, respectively, consistent with the results of the first-principles calculations. Moreover, the binding energy of about 75 meV per S atom was obtained, implying a weak interfacial interaction between the layered WS$_2$ and SiO$_2$ surface.

**2 Experimental details**

Atomically thin WS$_2$ layers were deposited on 300 nm SiO$_2$/Si substrates by using ultrahigh vacuum dc magnetron sputtering. High purity W target and S particles were used as the reaction sources. More growth details can be also found in our previous works [8,36,37]. S partial pressure was controlled to be $3.0 \times 10^{-5}$ millibars and the substrate temperature is set to be 680 °C. Ar pressure of $1.1 \times 10^{-3}$ millibars and a low sputtering power of only 8 W were adopted to achieve the thickness controllable growth. When growing for 30 s under above conditions, the $WS_2$ monolayer could be obtained. The surface image and thickness of as-grown monolayer $WS_2$ films were further taken by atomic force microscopy (AFM) using the Bruker Dimension Icon system. Raman and photoluminescence spectra were taken to confirm the thickness and crystal quality of $WS_2$ films and performed by a Thermo Scientific DXR microscope with a 514.5 nm laser. Atomic structure was observed by using the TEM JEOL 2100 system. XPS system (VG ESCALAB 220i-XL) was mainly adopted to investigate the interfacial properties of the layered $WS_2/SiO_2$ heterostructures. The relative sensitivity factors for W and S elements are respectively 9.8 and 1.67 in the XPS measurement. C 1s signal from the sample surface was used to correct the core-level binding energy.

**3 Results and discussion**

Figure 1(a) shows the AFM image of as-grown monolayer $WS_2$ film on the $SiO_2$ substrate. The cross sectional height of about 0.91 nm along with the white dashed line in Fig. 1(a) confirms the formation of the 1L $WS_2$ on the $SiO_2$ substrate. Raman spectrum of 1L $WS_2$ on $SiO_2$ wafer shows the $E_{2g}^1$ (in-plane) and $A_{1g}$ (out-of-plane) peaks at 357.0 and 418.8 cm$^{-1}$, respectively, as given in Fig. 1(c). The stronger intensity of $E_{2g}^1$ peak and the frequency difference of about 61.8 cm$^{-1}$ [38] further indicate the formation of the 1L $WS_2$. A strong exciton emission peak at 611 nm was obtained in the photoluminescence spectrum, as shown in Fig. 1(d), implying the high quality of the monolayer $WS_2$ film. The typical transmission electron microscopy (TEM) images, as shown in Fig. 1(e) and 1(f), further confirm the hexagonal 2H phase of the as-grown $WS_2$ layer. The periodic arrangement of W and S atoms were clearly observed, indicating the successful fabrication of the high-quality layered $WS_2$ film. The lattice constant of 2H-$WS_2$ crystal is 0.317 nm, similar with the reported values [8].

The core-level XPS spectra of W 4f and S 2p were taken from 1L and bulk $WS_2$/$SiO_2$ interfaces and shown in Fig. 2. During the XPS fitting, we followed the rule that the full width at half maximum (FWHM) of both components are comparable and the intensity ratios follow the expect quantum mechanically predicted ratio when XPS doublet have narrow separations. For W 4f spectrum of 1L $WS_2$, the peaks at 33.3 and 35.5 eV originate respectively from the 2H-$WS_2$ $W^{4+}$ $4f_{7/2}$ and $W^{4+}$ $4f_{5/2}$ orbitals [39]. For S 2p spectrum of 1L $WS_2$, the well resolved peaks at 163.0 and 164.2 eV originate respectively from the $S^{2-}$ $2p_{3/2}$ and $S^{2-}$ $2p_{1/2}$ orbitals. Based on the quantitative analysis of XPS peaks, the W/S ratio is further calculated to be about 1:2, revealing the correct stoichiometry and high quality of the as-grown layered $WS_2$ [40]. The peaks of W 4f spectrum at 36.5 and 38.3 eV may be $W^{6+}$ $4f_{7/2}$ and $W^{6+}$ $4f_{5/2}$ orbitals originated from the minor $WO_3$ contributions (~2%) [39]. Furthermore, C 1s signal from the sample surface was used to correct the core-level binding energy in this study, then the existence of the small amounts of $WO_3$ will not impact the determination of the band alignment of the $WS_2$/$SiO_2$ interfaces. Additionally, the W $5p_{3/2}$ peak overlapping with the $W^{6+}$ $4f_{5/2}$ is also observed at 38.9 eV. For bulk $WS_2$, the $W^{4+}$ $4f_{7/2}$, $W^{4+}$ $4f_{5/2}$, $S^{2-}$ $2p_{3/2}$ and $S^{2-}$ $2p_{1/2}$ peaks are located at 33.4, 35.6, 163.0, and 164.1 eV, respectively, similar with those of 1L WS$_2$. The $W^{6+}$ $4f_{7/2}$, $W^{6+}$ $4f_{5/2}$ and W $5p_{3/2}$ peaks are located at 36.4, 38.3 and 39.0 eV, respectively. On the whole, W 4f and S 2p spectra show similar distributions for 1L and bulk WS$_2$ grown on the SiO$_2$ substrates. It is noted that W 4f and S 2p spectra from bulk WS$_2$ show much stronger intensities than those from 1L WS$_2$, which further implies from another point of view the formation of the 1L and bulk WS$_2$ in our experiments. Above obtained W 4f and S 2p spectra agree well with the current literature values [8,39-41]. More characterization on our 2D WS$_2$ can be also found in our previous works [8,36,37].

XPS has been proved to be an efficient and noninvasive technique for acquiring the interfacial properties of the heterojunction systems. We first adopted the indirect method proposed by Kraut *et al.* [42] to investigate the band alignment at 1L WS$_2$/SiO$_2$ interface. The relative energy positions between the core and valence levels are assumed to be unaltered both in the bulk and at its interface. Then the valence band offset (VBO) can be calculated by

$$\Delta E_V = (E_{CL}^{bulk\ WS_2} - E_V^{bulk\ WS_2}) - (E_{CL}^{bulk\ SiO_2} - E_V^{bulk\ SiO_2}) - (E_{CL}^{interface\ WS_2} - E_{CL}^{interface\ SiO_2}), \quad (1)$$

where $(E_{CL}^{bulk\ WS_2} - E_V^{bulk\ WS_2})$ is the binding energy difference between the core-level $E_{CL}^{bulk\ WS_2}$ and valence band maximum (VBM) $E_V^{bulk\ WS_2}$ of bulk WS$_2$. $(E_{CL}^{bulk\ SiO_2} - E_V^{bulk\ SiO_2})$ is the binding energy difference between the core-level $E_{CL}^{bulk\ SiO_2}$ and VBM $E_V^{bulk\ SiO_2}$ of bulk SiO$_2$. $(E_{CL}^{interface\ WS_2} - E_{CL}^{interface\ SiO_2})$ is the core-level difference between upper 1L WS$_2$ and lower SiO$_2$ substrate. Fig. 3(a) and 3(b) show the core-level and valence band spectra, obtained respectively from bulk WS$_2$ and bulk SiO$_2$. Fig. 3(c) is the core-level spectra of S 2p and Si 2p obtained from the 1L WS$_2$/SiO$_2$ heterostructure. The VBM was derived by the intersection of the linear regressions of the leading edge of valence band and the baseline of its spectra. $E_{CL}$ for WS$_2$ and SiO$_2$ was taken to be the binding energy of S$^{2-}$ 2p$_{3/2}$ and Si 2p peaks, respectively. The binding energy differences $(E_{CL}^{bulk\ WS_2} - E_{V}^{bulk\ WS_2})$, $(E_{CL}^{bulk\ SiO_2} - E_{V}^{bulk\ SiO_2})$, and $(E_{CL}^{interface\ WS_2} - E_{CL}^{interface\ SiO_2})$ have been illustrated in Fig. 3(a)-3(c). Using Eq. (1), the valence band offset (VBO) of 1L WS$_2$/SiO$_2$ heterostructure was calculated to be 3.86 eV. Then, the conduction band offset (CBO) at 1L WS$_2$/SiO$_2$ interface can be given by

$$\Delta E_C = E_{gap}^{SiO_2} - E_{gap}^{1L\ WS_2} - \Delta E_V, \qquad (2)$$

where $E_{gap}^{SiO_2}$ and $E_{gap}^{1L\ WS_2}$ are respectively the bandgap of SiO$_2$ and 1L WS$_2$. Based on the O 1s loss energy spectrum, as shown in Fig. 3(d), the bandgap of SiO$_2$ is obtained to be 9.05 eV, consistent with the previous results [43]. Since the electronic bandgap of 2.38 eV was obtained for 1L WS$_2$ in the scanning tunneling spectroscopy (STS) measurement [16], the CBO at 1L WS$_2$/SiO$_2$ interface is then obtained to be 2.81 eV.

As a comparison, the direct method used by Santoni *et al*. [44] was further adopted to derive the band offsets of the WS$_2$/SiO$_2$ heterostructures. This method requires the alignment of the XPS spectra of bare SiO$_2$, 1L and bulk WS$_2$/SiO$_2$ to a common reference before comparing their relative VBM positions. Here, we used the Si 2p spectrum (104.50 eV) from the bare SiO$_2$ wafer as a calibration to align the valence bands of bare SiO$_2$, 1L WS$_2$/SiO$_2$ and bulk WS$_2$/SiO$_2$ samples [45,46], as shown in Fig. 4. Since that, $E_{CL}^{bulk\ SiO_2}$ will be equal to $E_{CL}^{interface\ SiO_2}$ in equation (1). Further substituting $E_{CL}^{bulk\ WS_2}$ and $E_{V}^{bulk\ WS_2}$ by $E_{CL}^{interface\ WS_2}$ and $E_{V}^{interface\ WS_2}$, respectively, equation (1) will change to

$$\Delta E_V = E_V^{bulk\ SiO_2} - E_V^{interface\ WS_2} . \qquad (3)$$

Therefore, the direct method could deduce both the band offsets at the 1L and bulk WS$_2$/SiO$_2$ interfaces as compared to the indirect method. The leading edge of valence band of clean SiO$_2$ wafer is obtained to be 5.88 eV. The Si 2p and valence band spectra taken from 1L WS$_2$/SiO$_2$ heterostructure were shown in Fig. 4(b). As compared to Fig. 4(a), some new electronic states at ~3.03 eV are clearly observed, which is owing to the formation of hybridized W 5d-S 3p states [47]. These hybridized W 5d-S 3p states indicates the formation of long-range ordered 2H-WS$_2$ film. The valence band edge of 1L WS$_2$/SiO$_2$ heterostructure is obtained to be 1.91 eV, as illustrated in Fig. 4(b). Further considering the edge of O 2p band (5.88 eV) from the SiO$_2$ wafer, the VBO of 3.97 eV at 1L WS$_2$/SiO$_2$ interface was obtained. Taking into account the same bandgap values of the 1L WS$_2$ (2.38 eV) and SiO$_2$ (9.05 eV) as the previous calculation yields the CBO value of 2.70 eV at 1L WS$_2$/SiO$_2$ interface. Therefore, the band offsets of the 1L WS$_2$/SiO$_2$ heterostructure obtained from the direct method are in agreement with those from the indirect Kraut's method, indicating the validity of the experimental XPS results. Furthermore, the VBO derived from the direct method agree well with the PEEM results of 4.0 eV [35]. However, there are still small differences between two methods, which might be owing to ignoring the quantum size effects in the 1L $WS_2$ sample by the indirect method [48].

Furthermore, the valence band spectrum of bulk $WS_2/SiO_2$ interface was taken and shown in Fig. 4(c), where the Si 2p spectrum from the $SiO_2$ substrate was also given as the reference. The electronic states from the hybridized W 5d-S 3p orbitals are developed at ~ 2.68 eV. The valence band edge is obtained to be 1.43 eV. Therefore, the formation of bulk $WS_2$ shifts the valence band edge by 0.48 eV towards the Fermi level in comparison with that of the 1L $WS_2/SiO_2$ interface. Then, the VBO value of 4.45 eV at the bulk $WS_2/SiO_2$ interface was obtained. When the bandgap values of bulk $WS_2$ (1.57 eV) and $SiO_2$ (9.05 eV) were employed [17], the CBO value of 3.03 eV at bulk $WS_2/SiO_2$ interface is then determined. Therefore, the VBO (CBO) value at bulk $WS_2/SiO_2$ interface is about 0.48 eV (0.33 eV) larger than that of 1L $WS_2/SiO_2$ interface, which will be further discussed based on the following first-principles calculations. In addition, we further collected the XPS spectrum at different positions of different samples, and the average VBOs of 1L and bulk $WS_2/SiO_2$ interfaces are about 4.10 and 4.49 eV, respectively. The consistent VBOs prove the accuracy of the experiments and the uniformity of the as-grown $WS_2$. It is known that $SiO_2$ is a good gate insulator for FETs due to its high surface or interfacial quality. Meanwhile, sufficient VBO and CBO at the interface of $SiO_2$/semiconductor are still necessary to act as barriers for both electron and hole injections to avoid gate leakage. Clearly, the VBO and CBO values of the $WS_2/SiO_2$ interfaces are sufficient (>1.0 eV) for the suppression of the leakage current, indicating the applicability of the $WS_2/SiO_2$ heterojunction for high performance FETs.

To further investigate the interfacial interaction in the $WS_2/SiO_2$ heterostructures, first-principles calculations were performed by using Vienna ab initio simulation package (VASP) [49]. The exchange-correlation function adopted in this calculation was the Perdew-Burke-Ernzerhof (PBE) function [50]. The van der Waals effect was included in the calculation by the DFT-D3 approach [51]. To obtain appropriate band alignment, Heyd-Scuseria-Ernzerhof (HSE06) calculations were performed to simulate the density of states (DOSs) of the $WS_2/SiO_2$ heterostructure [52]. Possible artificial dipole interactions were eliminated by the dipole correction [53]. For 1L $WS_2$, the optimized lattice constant and HSE06 bandgap are 3.19 Å and 2.37 eV, respectively. For bulk alpha-$SiO_2$, the optimized lattice constant and HSE06 bandgap are 5.02 Å and 7.53 eV, respectively. All the basic parameters obtained in this calculations agree well with the reported values [47,54].

To simulate experimentally relevant situations, the amorphous $SiO_2$ (a-$SiO_2$) bulk structure was further generated through the molecular dynamics (MD) simulations based on the VASP code following the same procedure as showed in detailed in literatures [55,56]. The $WS_2/SiO_2$ heterostructure was further simulated by constructing a repeated-slab model, including a-$SiO_2$, (3×3)-$WS_2$, and a 20 Å vacuum region. The optimized configuration of 1L $WS_2$/a-$SiO_2$ interface was shown in Fig.

5(a), in which the (3×3) WS$_2$ monolayer was located on the 4.53% compressed a-SiO$_2$ surface. The binding energy between 1L WS$_2$ and the a-SiO$_2$ surface is obtained to be only 75 meV per S atom, confirming the weak interaction at the 1L WS$_2$/SiO$_2$ interface. Moreover, 1L WS$_2$ almost keeps its intact hexagonal network. Our result is in agreement well with the case of the graphene weakly adsorbed on the SiO$_2$ surface without forming the covalent bond between them [57]. Furthermore, most recent research about the novel fluorescence aging behavior induced by strain release in the WS$_2$/SiO$_2$ heterostructure [58] also proved from another side the weak interaction at the WS$_2$/SiO$_2$ interface as demonstrated in this study.

Figure 5(b) presents the calculated total DOSs of 1L WS$_2$/SiO$_2$ interface and the partial DOSs of Si, O, W and S atoms. The valence bands of SiO$_2$ substrate are located below -3.17 eV, and the conduction bands are above 4.82 eV. The contributions to conduction band minimum (CBM) and VBM of 1L WS$_2$/SiO$_2$ heterostructure mainly come from W and S atoms of 1L WS$_2$. No surface or interfacial electronic states were found in the energy region from -3.17 to 4.82 eV, indicating the totally separated electronic states of the upper 1L WS$_2$ and lower SiO$_2$ substrate and negligible Fermi level pinning effect. The separated DOS distributions and negligible Fermi level pinning effect further imply the weak interfacial interaction at the 1L WS$_2$/SiO$_2$ interface. Furthermore, the VBO (3.17 eV) and CBO (2.58 eV) at 1L WS$_2$/SiO$_2$ interface could be identified and illustrated in Fig. 5(b). Additionally, it should be noted that the weak interaction also shows a tiny modification of the bandgaps of the 1L $WS_2$ (2.24 eV) and the $SiO_2$ substrate (7.99 eV) obtained in the $WS_2/SiO_2$ heterostructure, similar with the observation in the graphene/$SiO_2$ heterostructure [57].

Above XPS results show that the VBO (CBO) value at bulk $WS_2/SiO_2$ interface is about 0.48 eV (0.33 eV) larger than that of 1L $WS_2/SiO_2$ interface. To illuminate the experimental findings, partial DOSs of W 5d and S 3p calculated from 1L and bulk $WS_2$ were presented in Fig. 5(c). The low-energy electronic states were taken as references to align the partial DOSs. Compared with the W 5d and S 3p states of 1L $WS_2$, some new electronic states emerge around the VBM and CBM and further reduce the bandgap of bulk $WS_2$. According to the previous theoretical calculation [47], these electronic states originate from the splitting of valence and conduction band edges due to the interlayer orbital coupling in bulk $WS_2$. Then, the band offsets for the bulk $WS_2/SiO_2$ heterostructure will be enlarged correspondingly. The shift-up energy around the VBM is observed to be about 0.47 eV, which agrees well with the above XPS result of 0.48 eV. By using the photoemission electron microscopy, Keyshar *et al*. also obtained the shift-up energy around the VBM of about 0.48 eV between 1L and bulk $WS_2$ [59]. Thus, the VBO at bulk $WS_2/SiO_2$ interface can be calculated to be 3.64 eV. Since the bandgap of 1.52 eV for bulk $WS_2$ is obtained in this calculation, the CBO value of 2.83 eV could be obtained at bulk $WS_2/SiO_2$ interface.

We also constructed another configuration of 1L WS$_2$/O-terminated (2×2) α-SiO$_2$(0001) heterostructure, as shown in Fig. 6(a). The binding energy between 1L WS$_2$ and the compressed O-terminated (2×2) α-SiO$_2$(0001) surface is obtained to be 165 meV per S atom. The interlayer space between 1L WS$_2$ and SiO$_2$ substrate is 3.09 Å. Small binding energy and wide interlayer space confirm that the interaction at the 1L WS$_2$/SiO$_2$ interface is week. The calculated partial DOSs of W, S, Si, and O atoms [as shown in Fig. 6(b)] also exhibit the totally separated electronic states of the upper 1L WS$_2$ and lower SiO$_2$ substrate and negligible Fermi level pinning effect. And the VBO (3.52 eV) and CBO (2.22 eV) of this configuration could be identified, which agrees well with the theoretical results of the 1L WS$_2$/a-SiO$_2$ heterostructure.

Figures 7(a) and 7(b) show the experimental band alignments of the 1L and bulk WS$_2$/SiO$_2$ heterostructures derived using the indirect and direct method. For comparison, theoretical band alignments calculated from the 1L WS$_2$/a-SiO$_2$ heterostructure were also shown in Fig. 7(c). It can be seen that WS$_2$/SiO$_2$ heterostructures have a Type I alignment, where the conduction and valence band edge of 1L and bulk WS$_2$ films are both located within the bandgap of the SiO$_2$ substrate. For 1L WS$_2$/SiO$_2$ heterostructure, the results from both experimental indirect and direct methods exhibit the similar distributions. On the whole, the theoretical VBOs and CBOs are consistent with the experimental XPS band offsets, except that the theoretical VBOs and CBOs are about 0.8 and 0.2 eV smaller than the experimental results, respectively. Since the bandgap difference between the experimental (9.05 eV) and theoretical (7.99 eV) is 1.06 eV, the numerical discrepancies between the theoretical and experimental VBOs and CBOs might be just due to the narrower theoretical bandgap of $SiO_2$. Thus, the experimental band offsets from XPS are agreement well with the first-principles calculations, which also confirms the weak interaction at the $WS_2/SiO_2$ interface. The sufficient VBOs and CBOs and weak interaction reveal that the integration strategy of 2D $WS_2/SiO_2$ heterostructure still play an important role in fabrication of the next-generation electronic and optoelectronic devices.

**4 Conclusions**

In summary, the interfacial properties of 2D layered $WS_2$ grown on $SiO_2/Si$ wafers were investigated using XPS and first-principles calculations. By using the experimental direct (indirect) method, the VBO value of 3.97 eV (3.86 eV) was obtained at 1L $WS_2/SiO_2$ interface, while the CBO was obtained to be 2.70 eV (2.81 eV). Furthermore, the VBO (CBO) value at bulk $WS_2/SiO_2$ interface is about 0.48 eV (0.33 eV) larger than that of 1L $WS_2/SiO_2$ interface. Therefore, the $WS_2/SiO_2$ heterostructures have a Type I energy-band alignment. The band offsets obtained experimentally and theoretically are consistent except the narrower theoretical bandgap of $SiO_2$. Moreover, the binding energy of 75 meV per S atom and the totally separated partial DOSs demonstrate a weak interaction and negligible Fermi level pinning effect between $WS_2$ monolayer and $SiO_2$ surface. Our experimental and theoretical results ensure the practical applications of 2D WS$_2$/SiO$_2$ heterostructures in next-generation electronic and optoelectronic devices.

**Acknowledgements**

This work was supported by the National Natural Science Foundation of China (Grant No. 11804115), the Foundation from Department of Science and Technology of Fujian Province (Grant Nos. 2019L3008, 2020J01704, 2021J01863, 2021J05171), the Foundation from Department of Education of Fujian Province (Grant No. JT180261) and the Scientific Research Foundation from Jimei University (Grant Nos. ZC2018007, ZQ2019008, ZP2020066 and ZP2020065).

**References**

1. K. F. Mak, C. Lee, J. Hone, J. Shan, and T. F. Heinz, Atomically thin MoS$_2$: a new direct-gap semiconductor, Phys. Rev. Lett. 105(13), 136805 (2010)

2. G. Eda, H. Yamaguchi, D. Voiry, T. Fujita, M. W. Chen, and M. Chhowalla, Photoluminescence from chemically exfoliated MoS$_2$, Nano Lett. 11(12), 5111 (2011)

3. J. N. Coleman, M. Lotya, A. O'Neill, S. D. Bergin, P. J. King, U. Khan, K. Young, A. Gaucher, S. De, R. J. Smith, I. V. Shvets, S. K. Arora, G. Stanton, H. Y. Kim, K. Lee, G. T. Kim, G. S. Duesberg, T. Hallam, J. J. Boland, J. J. Wang, J. F. Donegan, J. C. Grunlan, G. Moriarty, A. Shmeliov, R. J. Nicholls, J. M. Perkins, E. M. Grieveson, K. Theuwissen, D. W. McComb, P. D. Nellist, and V. Nicolosi, Two-dimensional nanosheets produced by liquid exfoliation of layered materials, Science 331 (6017), 568 (2011)

4. Y. J. Zhan, Z. Liu, S. Najmaei, P. M. Ajayan, and J. Lou, Large-area vapor-phase growth and characterization of $MoS_2$ atomic layers on a $SiO_2$ substrate, Small 8(7), 966 (2012)

5. Y. H. Lee, X. Q. Zhang, W. J. Zhang, M. T. Chang, C. T. Lin, K. D. Chang, Y. C. Yu, J. T. W. Wang, C. S. Chang, L. J. Li, and T. W. Lin, Synthesis of large-area $MoS_2$ atomic layers with chemical vapor deposition, Adv. Mater. 24(17), 2320 (2012)

6. P. Y. Liu, T. Luo, J. Xing, H. Xu, H. Y. Hao, H. Liu, and J. J. Dong, Large-area $WS_2$ film with big single domains grown by chemical vapor deposition, Nanoscale Res. Lett. 12, 558 (2017)

7. A. L. Elias, N. Perea-Lopez, A. Castro-Beltran, A. Berkdemir, R. T. Lv, S. M. Feng, A. D. Long, T. Hayashi, Y. A. Kim, M. Endo, H. R. Gutierrez, N. R. Pradhan, L. Balicas, T. E. Mallouk, F. Lopez-Urias, H. Terrones, and M. Terrones, Controlled synthesis and transfer of large-area $WS_2$ sheets: from single layer to few layers, Acs Nano. 7(6), 5235 (2013)

8. H. L. Zhu, C. J. Zhou, B. S. Tang, W. F. Yang, J. W. Chai, W. L. Tay, H. Gong, J. S. Pan, W. D. Zou, S. J. Wang, and D. Z. Chi, Band alignment of 2D $WS_2$/$HfO_2$ interfaces from x-ray photoelectron spectroscopy and first-principles calculations, Appl. Phys. Lett. 112 (17), 171604 (2018)

9. M. X. Ye, D. Y. Zhang, and Y. K. Yap, Recent advances in electronic and optoelectronic devices based on two-dimensional transition metal dichalcogenides, Electronics 6(2), 43 (2017)

10. C. X. Cong, J. Z. Shang, Y. L. Wang, and T. Yu, Optical properties of 2D semiconductor $WS_2$, Adv. Opt. Mater. 6(1), 1700767 (2018)

11. P. J. Schuck, W. Bao, and N. J. Borys, A polarizing situation: taking an in-plane perspective for next-generation near-field studies, Front. Phys. 11(2), 117804 (2016)

12. Z. C. Zhou, F. Y. Yang, S. Wang, L. Wang, X. F. Wang, C. Wang, Y. Xie, and Q. Liu, Emerging of two-dimensional materials in novel memristor, Front. Phys. 17(2), 23204 (2021)

13. H. L. Zhu, W. H. Yang, Y. P. Wu, W. Lin, J. Y. Kang, and C. J. Zhou, Au and Ti induced charge redistributions on monolayer $WS_2$, Chin. Phys. B 24(7), 077301 (2015)

14. T. LaMountain, E. J. Lenferink, Y. J. Chen, T. K. Stanev, and N. P. Stern, Environmental engineering of transition metal dichalcogenide optoelectronics, Front. Phys. 13(4), 138114 (2019)

15. G. Luo, Z. Z. Zhang, H. O. Li, X. X. Song, G. W. Deng, G. Cao, M. Xiao, and G. P. Guo, Quantum dot behavior in transition metal dichalcogenides nanostructures, Front. Phys. 12(4), 128502 (2017)

16. H. M. Hill, A. F. Rigosi, K. T. Rim, G. W. Flynn, and T. F. Heinz, Band alignment in $MoS_2/WS_2$ transition metal dichalcogenide heterostructures probed by scanning tunneling microscopy and spectroscopy, Nano lett. 16(8), 4831 (2016)

17. Y. Z. Guo and J. Robertson, Band engineering in transition metal dichalcogenides: stacked versus lateral heterostructures, Appl. Phys. Lett. 108(23), 233104 (2016)

18. J. Jadczak, J. Kutrowska-Girzycka, P. Kapuscinski, Y. S. Huang, A. Wojs, and L. Bryja, Probing of free and localized excitons and trions in atomically thin $WSe_2$, $WS_2$, $MoSe_2$ and $MoS_2$ in photoluminescence and reflectivity experiments, Nanotechnology 28(39), 395702 (2017)

19. P. K. Nayak, F. C. Lin, C. H. Yeh, J. S. Huang, and P. W. Chiu, Robust room temperature valley polarization in monolayer and bilayer $WS_2$, Nanoscale 8(11), 6035 (2016)

20. N. Ubrig, S. Jo, M. Philippi, D. Costanzo, H. Berger, A. B. Kuzmenko, and A. F. Morpurgo, Microscopic origin of the valley hall effect in transition metal dichalcogenides revealed by wavelength-dependent mapping, Nano Lett. 17(9), 5719 (2017)

21. M. Van der Donck, M. Zarenia, and F. M. Peeters, Strong valley Zeeman effect of dark excitons in monolayer transition metal dichalcogenides in a tilted magnetic field, Phys. Rev. B 97(8), 081109 (2018)

22. M. W. Iqbal, M. Z. Iqbal, M. F. Khan, M. A. Kamran, A. Majid, T. Alharbi, and J. Eom, Tailoring the electrical and photo-electrical properties of a $WS_2$ field effect transistor by selective n-type chemical doping, RSC Adv. 6(29), 24675 (2016)

23. N. J. Huo, S. X. Yang, Z. M. Wei, S. S. Li, J. B. Xia, and J. B. Li, Photoresponsive and gas sensing field-effect transistors based on multilayer $WS_2$ nanoflakes, Sci. Rep. 4, 5209 (2014)

24. D. Akinwande, N. Petrone, and J. Hone, Two-dimensional flexible nanoelectronics, Nat. Commun. 5, 5678 (2014)

25. Y. Wang, D. Z. Kong, S. Z. Huang, Y. M. Shi, M. Ding, Y. V. Lim, T. T. Xu, F. M. Chen, X. J. Li, and H. Y. Yang, 3D carbon foam-supported $WS_2$ nanosheets for cable-shaped flexible sodium ion batteries, J. Mater. Chem. A 6(23), 10813 (2018)

26. C. Y. Lan, Z. Y. Zhou, Z. F. Zhou, C. Li, L. Shu, L. F. Shen, D. P. Li, R. T. Dong, S. P. Yip, and J. Ho, Wafer-scale synthesis of monolayer $WS_2$ for high-performance flexible photodetectors by enhanced chemical vapor deposition, Nano Res. 11(6), 3371 (2018)

27. C. Ouyang, Y. X. Chen, Z. Y. Qin, D. W. Zeng, J. Zhang, H. Wang, and C. S. Xie, Two-dimensional $WS_2$-based nanosheets modified by Pt quantum dots for enhanced room-temperature $NH_3$ sensing properties, Appl. Surf. Sci. 455, 45 (2018)

28. G. A. Asres, J. J. Baldovi, A. Dombovari, T. Jarvinen, G. S. Lorite, M. Mohl, A. Shchukarev, A. P. Paz, L. D. Xian, J. P. Mikkola, A. L. Spetz, H. Jantunen, A. Rubio, and K. Kordas, Ultrasensitive H$_2$S gas sensors based on p-type WS$_2$ hybrid materials, Appl. Nano Res. 11(8), 4215 (2018)

29. Q. H. Xu, Y. T. Lu, S. Y. Zhao, N. Hu, Y. W. Jiang, H. Li, Y. Wang, H. Q. Gao, Y. Li, M. Yuan, L. Chu, J. H. Li, and Y. N. Xie, A wind vector detecting system based on triboelectric and photoelectric sensors for simultaneously monitoring wind speed and direction, Nano Energy 89, 106382 (2021)

30. Q. H. Xu, Y. S. Fang, B. Q. S. Jing, N. Hu, K. Lin, Y. F. Pan, L. Xu, H. Q. Gao, M. Yuan, L. Chu, Y. W. Ma, Y. N. Xie, J. Chen, and L. H. Wang, A portable triboelectric spirometer for wireless pulmonary function monitoring, Biosens. Bioelectron. 187, 113329 (2021)

31. Z. Z. Yan, Z. H. Jiang, J. P. Lu, and Z. H. Ni, Interfacial charge transfer in WS$_2$ monolayer/CsPbBr$_3$ microplate heterostructure, Front. Phys. 13(4), 138115 (2018)

32. W. J. Yin, X. L. Zeng, B. Wen, Q. X. Ge, Y. Xu, G. Teobaldi, and L. M. Liu, The unique carrier mobility of Janus MoSSe/GaN heterostructures, Front. Phys. 16(3), 33501 (2020)

33. H. Wang, D. L. Ren, C. Lu, and X. B. Yan, Investigation of multilayer WS$_2$ flakes as charge trapping stack layers in non-volatile memories, Appl. Phys. Lett. 112(23), 231903 (2018)

34. O. Zheliuk, J. M. Lu, J. Yang, and J. T. Ye, Monolayer superconductivity in WS$_2$, Phys. Status Solidi RRL 11(9), 1700245 (2017)

35. S. Ulstrup, R. J. Koch, D. Schwarz, K. M. McCreary, B. T. Jonker, S. Singh, A. Bostwick, E. Rotenberg, C. Jozwiak, and J. Katoch, Imaging microscopic electronic contrasts at the interface of single-layer WS$_2$ with oxide and boron nitride substrates, Appl. Phys. Lett. 114(15), 151601 (2019)

36. W. F. Yang, H. Kawai, M. Bosman, B. S. Tang, J. W. Chai, W. L. Tay, J. Yang, H. L. Seng, H. L. Zhu, H. Gong, H. F. Liu, K. E. J. Goh, S. J. Wang, and D. Z. Chi, Interlayer interactions in 2D WS$_2$/MoS$_2$ heterostructures monolithically grown by in situ physical vapor deposition, Nanoscale 10(48), 22927 (2018)

37. B. S. Tang, Z. G. Yu, L. Huang, J. W. Chai, S. L. Wong, J. Deng, W. F. Yang, H. Gong, S. J. Wang, K. W. Ang, Y. W. Zhang, and D. Z. Chi, Direct n- to p-type channel conversion in monolayer/few-layer WS$_2$ field-effect transistors by atomic nitrogen treatment, ACS Nano 12(3), 2506 (2018)

38. H. R. Gutierrez, N. Perea-Lopez, A. L. Elias, A. Berkdemir, B. Wang, R. Lv, F. Lopez-Urias, V. H. Crespi, H. Terrones, and M. Terrones, Extraordinary room-temperature photoluminescence in triangular WS$_2$ monolayers, Nano Lett. 13(8), 3447 (2013)

39. L. Yang, X. B. Zhu, S. J. Xiong, X. L. Wu, Y. Shan, and P. K. Chu, Synergistic WO$_3$·2H$_2$O Nanoplates/WS$_2$ hybrid catalysts for high-efficiency hydrogen evolution, ACS Appl. Mater. Inter. 8(22), 13966 (2016)

40. R. Bhandavat, L. David, and G. Singh, Synthesis of surface-functionalized $WS_2$ nanosheets and performance as Li-ion battery anodes, J. Phys. Chem. Lett. 3(11), 1523 (2012)

41. S. Cadot, O. Renault, D. Rouchon, D. Mariolle, E. Nolot, C. Thieuleux, L. Veyre, H. Okuno, F. Martin, and E. A. Quadrelli, Low-temperature and scalable CVD route to $WS_2$ monolayers on $SiO_2$/Si substrates, J. Vac. Sci. Technol. A 35(6), 061502 (2017)

42. E. A. Kraut, R. W. Grant, J. R. Waldrop, and S. P. Kowalczyk, Precise determination of the valence-band edge in x-ray photoemission spectra-application to measurement of semiconductor interface potentials, Phys. Rev. Lett. 44(24), 1620 (1980)

43. J. G. Tao, J. W. Chai, Z. Zhang, J. S. Pan, and S. J. Wang, The energy-band alignment at molybdenum disulphide and high-k dielectrics interfaces, Appl. Phys. Lett. 104(23), 232110 (2014)

44. A. Santoni, F. Biccari, C. Malerba, M. Valentini, R. Chierchia, and A. Mittiga, Valence band offset at the CdS/$Cu_2ZnSnS_4$ interface probed by x-ray photoelectron spectroscopy, J. Phys. D Appl. Phys. 46(17), 175101 (2013)

45. F. J. Grunthaner, B. F. Lewis, N. Zamini, J. Maserjian, and A. Madhukar, XPS studies of structure-induced radiation effects at the Si/$SiO_2$ interface, IEEE T. Nucl. Sci. 27(6), 1640 (1980)

46. J. Zhang, S. H. Wei, X. L. Wang, J. J. Xiang, and W. W. Wang, Experimental estimation of charge neutrality level of $SiO_2$, Appl. Surf. Sci. 422, 690 (2017)

47. H. L. Zhu, C. J. Zhou, X. J. Huang, X. L. Wang, H. Z. Xu, Y. Lin, W. H. Yang, Y. P. Wu, W. Lin, and F. Guo, Evolution of band structures in $MoS_2$-based homo- and heterobilayers, J. Phys. D Appl. Phys. 49(6), 065304 (2016)

48. Y. K. Lin, R. S. Chen, T. C. Chou, Y. H. Lee, Y. F. Chen, K. H. Chen, and L. C. Chen, Thickness-dependent binding energy shift in few-layer $MoS_2$ grown by chemical vapor deposition, ACS Appl. Mater. Inter. 8(34), 22637 (2016)

49. G. Kresse and J. Hafner, Ab initio molecular-dynamics for liquid-metals, Phys. Rev. B 47(1), 558 (1993)

50. J. P. Perdew, K. Burke, and M. Ernzerhof, Generalized gradient approximation made simple, Phys. Rev. Lett. 77(18), 3865 (1996)

51. S. Grimme, J. Antony, S. Ehrlich, and H. Krieg, A consistent and accurate ab initio parametrization of density functional dispersion correction (DFT-D) for the 94 elements H-Pu, J. Chem. Phys. 132(15), 154104 (2010)

52. J. Heyd, G. E. Scuseria, and M. Ernzerhof, Hybrid functionals based on a screened Coulomb potential, J. Chem. Phys. 118(18), 8207 (2003)

53. L. Bengtsson, Dipole correction for surface supercell calculations, Phys. Rev. B 59(19), 12301 (1999)

54. O. I. Malyi, V. V. Kulish, and C. Persson, In search of new reconstructions of (001) alpha-quartz surface: a first principles study, RSC Adv. 4(98), 55599 (2014)

55. W. L. Scopel, A. J. R. da Silva, and A. Fazzio, Amorphous $HfO_2$ and $Hf_{1-x}Si_xO$ via a melt-and-quench scheme using ab initio molecular dynamics, Phys. Rev. B 77(17), 172101 (2008)

56. W. L. Scopel, R. H. Miwa, T. M. Schmidt, and P. Venezuela, $MoS_2$ on an amorphous $HfO_2$ surface: An ab initio investigation, J. Appl. Phys. 117(19), 194303 (2015)

57. N. T. Cuong, M. Otani, and S. Okada, Semiconducting electronic property of graphene adsorbed on (0001) surfaces of $SiO_2$, Phys. Rev. Lett. 106(10), 106801 (2011)

58. S. H. Feng, R. L. Yang, Z. Y. Jia, J. Y. Xiang, F. S. Wen, C. P. Mu, A. M. Nie, Z. S. Zhao, B. Xu, C. G. Tao, Y. J. Tian, and Z. Y. Liu, Strain release induced novel fluorescence variation in CVD-grown monolayer $WS_2$ crystals, ACS Appl. Mater. Inter. 9(39), 34071 (2017)

59. K. Keyshar, M. Berg, X. Zhang, R. Vajtai, G. Gupta, C. K. Chan, T. E. Beechem, P. M. Ajayan, A. D. Mohite, and T. Ohta, Experimental determination of the ionization energies of $MoSe_2$, $WS_2$, and $MoS_2$ on $SiO_2$ using photoemission electron microscopy, ACS Nano 11(8), 8223 (2017)

**Figures and Figure Captions**

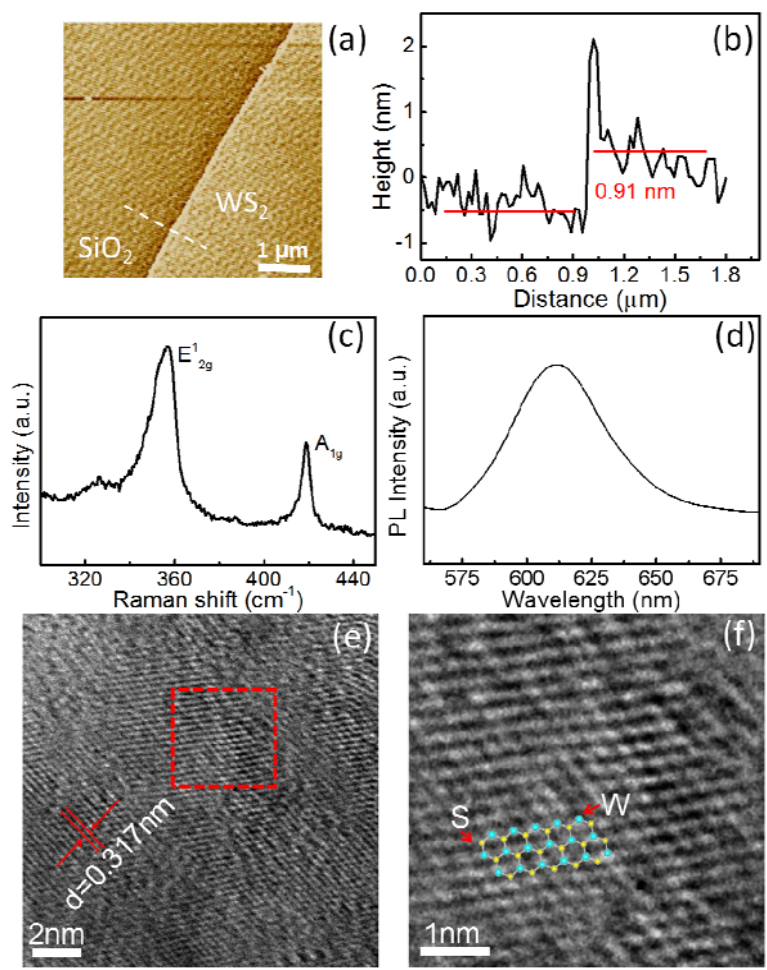

**Fig. 1** (a) AFM image of the monolayer WS$_2$ film (5×5 μm). (b) AFM height profile along with the white dashed line in (a). (c) Raman spectrum of WS$_2$ monolayer grown on SiO$_2$ wafer. (d) Photoluminescence of WS$_2$ monolayer. (e) Typical high-resolution TEM image of the WS$_2$ film. (f) Enlarged TEM image of a specific area illustrated by a red square in (e). The hexagonal atomic structures consist of W and S atoms are presented.

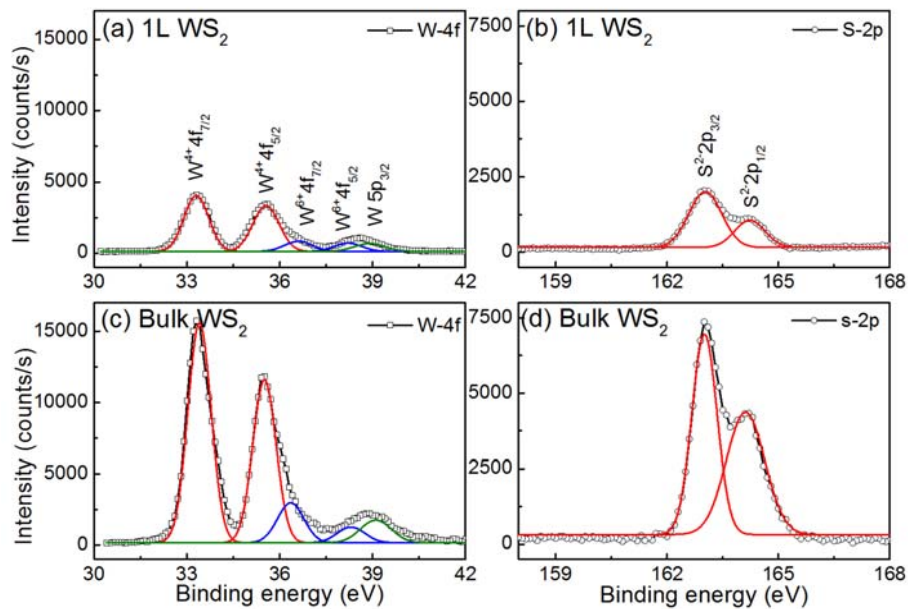

**Fig. 2** XPS W 4f and S 2p spectra of 1L and bulk $WS_2$ deposited on the $SiO_2$ substrates.

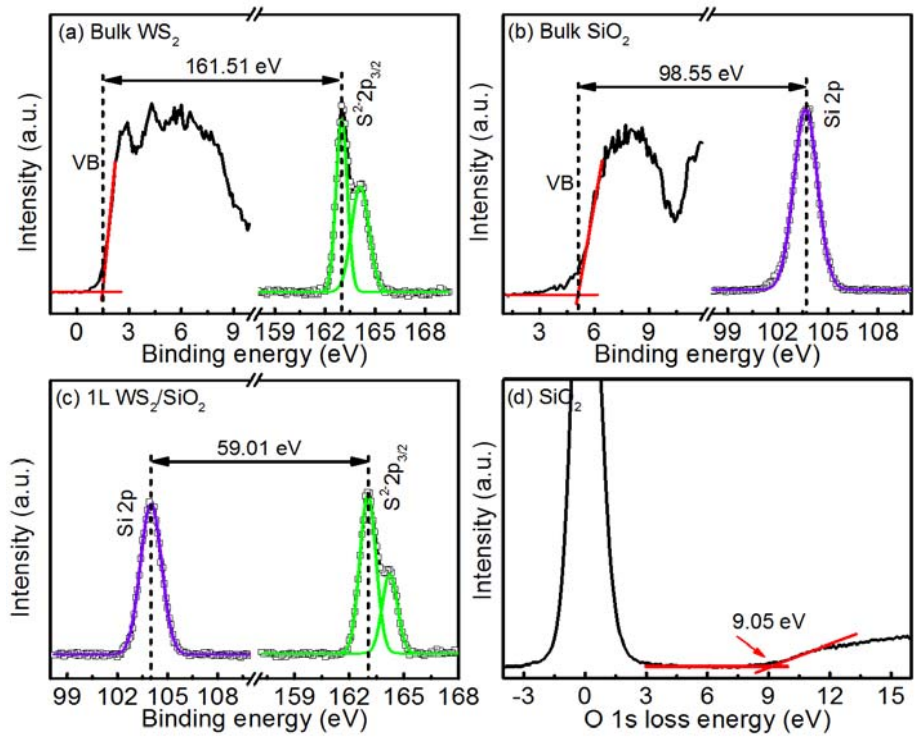

**Fig. 3** The core-level and valence band spectra for (a) bulk $WS_2$ and (b) bulk $SiO_2$ samples. (c) The core-level spectra of S 2p and Si 2p obtained for the 1L $WS_2$/$SiO_2$ heterostructure. (d) The bandgap of $SiO_2$ is measured to be 9.05 eV using O 1s loss energy spectrum.

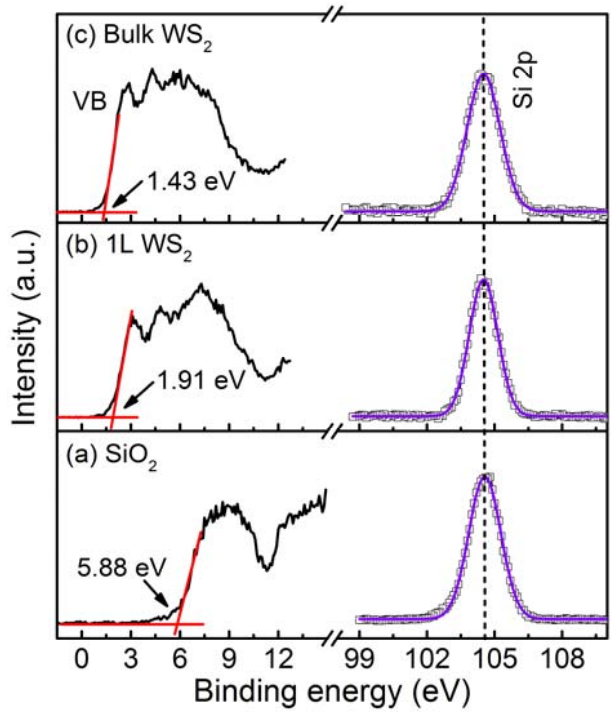

**Fig. 4** Si 2p and valence band spectra of (a) bare $SiO_2$ substrate, (b) 1L, and (c) bulk $WS_2$ deposited on $SiO_2$ substrates.

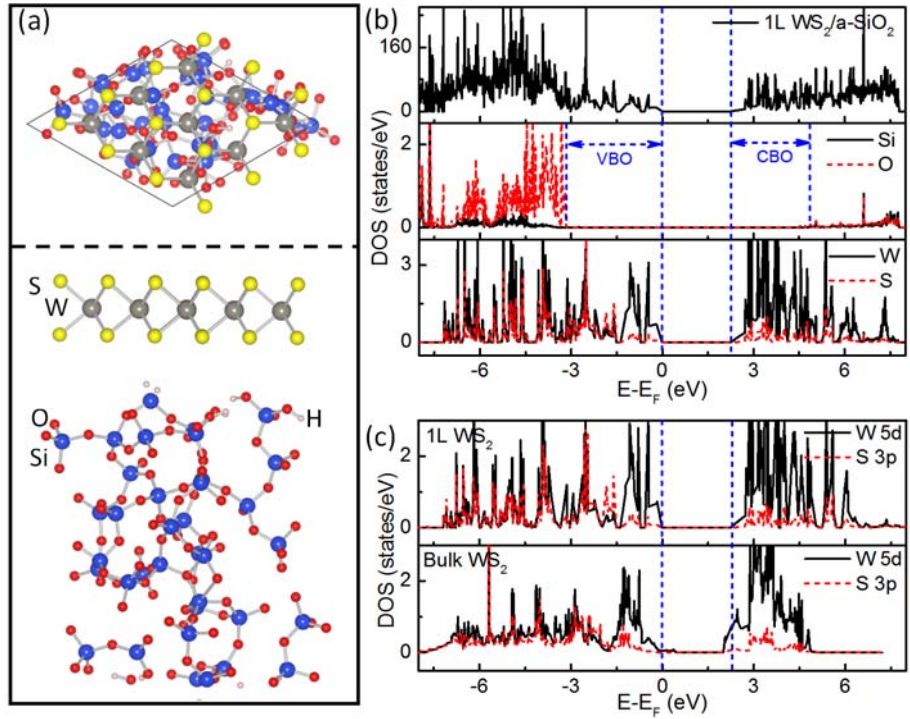

**Fig. 5** (a) Top and side view of the optimized atomic configuration of 1L $WS_2$/a-$SiO_2$ heterostructure. (b) The calculated total DOSs of 1L $WS_2$/a-$SiO_2$ interface and partial DOSs of Si, O , W, and S atoms based on the HSE06 method. The Fermi level is shifted to 0 eV. (c) Partial DOSs of W 5d and S 3p orbitals of 1L and bulk $WS_2$. The low-energy electronic states were taken as references to align the partial DOSs.

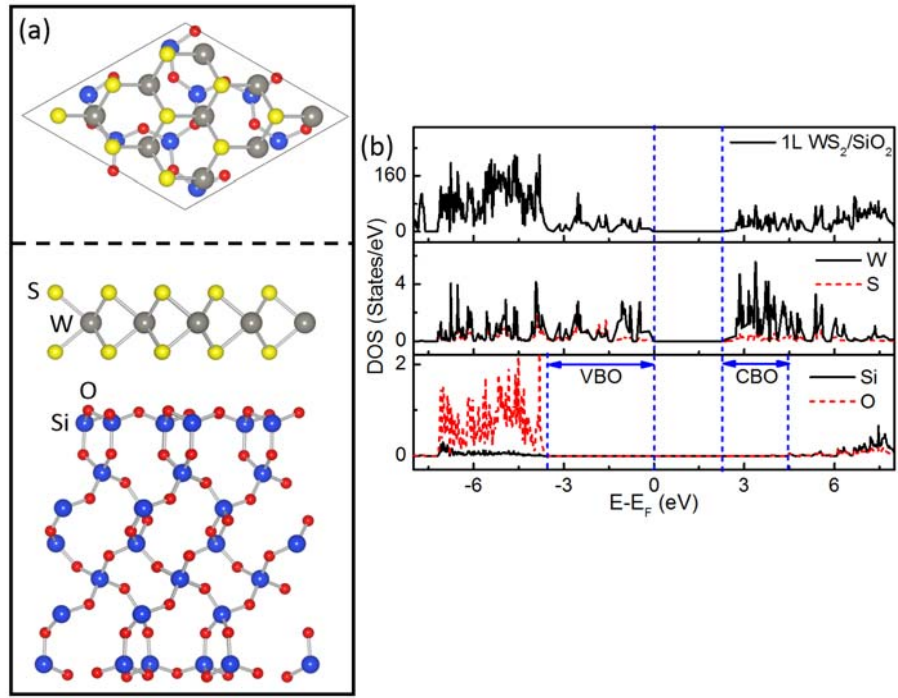

**Fig. 6** (a) Top and side view of the optimized atomic configuration of 1L WS$_2$/O-terminated (2×2) α-SiO$_2$(0001) heterostructure. (b) The calculated total DOSs of 1L WS$_2$/O-terminated (2×2) α-SiO$_2$(0001) interface and partial DOSs of W, S, Si, and O atoms based on the HSE06 method. The Fermi level is shifted to 0 eV.

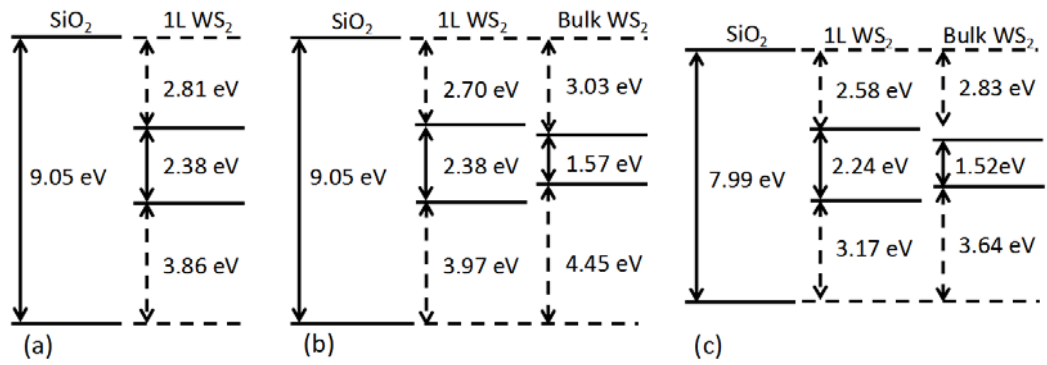

**Fig. 7** Experimental band alignments of 1L and bulk WS$_2$/SiO$_2$ heterostructures derived using the indirect (a) and direct (b) method. (c) Theoretical band alignments of 1L and bulk WS$_2$/a-SiO$_2$ heterostructures.